\documentclass[12pt]{article}
\newcommand{\beq}{\begin{equation}}
\newcommand{\eeq}{\end{equation}}
\newcommand{\beqa}{\begin{eqnarray}}
\newcommand{\eeqa}{\end{eqnarray}}
\newcommand{\beqar}{\begin{eqnarray*}}
\newcommand{\eeqar}{\end{eqnarray*}}

\newcommand{\eps}{\epsilon}

\newcommand{\inn}{\!\cdot\!}

\newcommand{\z}{\zeta}

\newcommand{\eg}{{\it e.g.,}\ }
\newcommand{\ie}{{\it i.e.,}\ }
\newcommand{\labell}[1]{\label{#1}} 
\newcommand{\reef}[1]{(\ref{#1})}
\newcommand\prt{\partial}

\newcommand\cF{{\cal F}}
\newcommand\cN{{\cal N}}

\newcommand\cA{{\cal A}}

\newcommand\cK{{\cal K}}
\newcommand\cM{{\cal M}}

\newcommand\cB{{\cal B}}

\newcommand\Tr{{\rm Tr}}

\begin{document}

\vspace*{1cm}

\begin{center}
{\bf \Large  On S-duality of D$_3$-brane S-matrix }

\vspace*{1cm}

{Mohammad R. Garousi\footnote{garousi@ferdowsi.um.ac.ir} }\\
\vspace*{1cm}
{ Department of Physics, Ferdowsi University of Mashhad,\\ P.O. Box 1436, Mashhad, Iran}
\\
\vspace{2cm}

\end{center}

\begin{abstract}
\baselineskip=18pt

There is a conjecture in the literature that indicates the tree-level S-matrix elements of graviton become symmetric under the $SL(2,Z)$ transformation after including the loops and the nonperturbative effects. 
Using the Ward identity corresponding to the global S-duality transformations,  this conjecture can be extended to other S-matrix elements as well. While the $SL(2,Z)$ transformation on the background dilaton is nonlinear,  the Ward identity dictates   the S-duality transformations on external states should be  the linearized $SL(2,R)$ transformations. We examine in details various S-matrix elements involving massless closed string and/or open string  vertex operators  on  the world volume of D$_3$-brane in favor of this conjecture.

\end{abstract}
Keywords: S-duality, S-matrix, D$_3$-brane

\vfill
\setcounter{page}{0}
\setcounter{footnote}{0}
\newpage


\section{Introduction } \label{intro}
It is known that the  type II superstring theory  is invariant under  T-duality \cite{Kikkawa:1984cp,TB,Giveon:1994fu,Alvarez:1994dn,Becker:2007zj}  and  S-duality \cite{Font:1990gx,Sen:1994fa,Rey:1989xj,Sen:1994yi,Schwarz:1993cr,Hull:1994ys,Becker:2007zj}. At the classical level, these dualities appear in  the equations of motion and in their solutions 
 \cite{Hassan:1991mq,Cvetic:1995bj,Breckenridge:1996tt,Costa:1996zd}. At the quantum level, these dualities should appear in the S-matrix elements. The  contact terms of the sphere-level S-matrix elements of graviton are speculated to be invariant under the S-duality after including the loops and the nonperturbative effects \cite{Green:1997tv} - \cite{Basu:2007ck}. This idea has been  extended to the S-matrix elements on the world volume of D$_3$-branes as well \cite{ Bachas:1999um,Basu:2008gt,Garousi:2011fc}. 
The S-duality  may also be used  to find the S-matrix elements on the world volume of F$_1$-string/NS$_5$-brane from the corresponding  S-matrix elements on the world volume of D$_1$-string/D$_5$-brane  \cite{Garousi:2011we}. 

The S-duality holds order by order in $\alpha'$ and is nonperturbative in the string loop expansion \cite{Becker:2007zj}.  So in order to study the S-duality of a S-matrix element, one has to first $\alpha'$-expand the amplitude in the Einstein frame and then study its S-duality at each order of $\alpha'$. Let us consider the  disk-level S-matrix element of two gravitons on the world volume of D$_3$-branes whose S-duality  has been studied in \cite{Bachas:1999um}. The leading $\alpha'$-order terms of this amplitude    
are invariant under the S-duality because the graviton in the Einstein frame is invariant.  On the other hand, the  non-leading order terms  which include the background dilaton factors as well, are speculated to become symmetric after including the loops and the non-perturbative effects \cite{Green:1997tv}. Using the fact that the S-matrix elements should satisfy the Ward identity corresponding to the global S-duality transformations, the above proposal has been  extended to  all   S-matrix elements in \cite{Garousi:2011we}. The S-dual Ward identity indicates that  an $n$-point function must transform  to an $n$-point function under the S-duality. Hence, in order to study the S-dual Ward identity of a S-matrix element one has to consider the linearized  $SL(2,R)$ transformations on the external states \cite{Garousi:2011we}. More specifically, the S-matrix element of  $n$-closed+$m$-open strings must transform to the S-matrix element of  $n$-closed+$m$-open strings, hence, the S-duality transformations of the external closed strings and the external open strings must be separately linear. The transformation of the background fields, however, should be the nonlinear S-duality transformation. In general, imposing the invariance of the S-matrix element under this later transformation, requires one to include the loops and the nonperturbative contributions to the tree-level S-matrix element.

All  S-matrix elements of  two massless RR and/or NSNS states \cite{Garousi:1996ad,Hashimoto:1996bf} on the world volume of D$_3$-brane have been analyzed in  \cite{Garousi:2011fc,Garousi:2011we} in favor of this proposal. 
In this paper we would like to test this proposal by examining in details  the S-matrix elements which involve the  open string states as well. In particular, using  the linear S-duality transformation of the gauge field \cite{Gibbons:1995ap}, we will show that the leading $\alpha'$-order terms of the S-matrix elements combine into linear S-duality invariant multiplets. The background dilaton factors in the nonleading order terms which are not invariant under the S-duality, may then be extending to the  $SL(2,Z)$ invariant functions after including the loops and the nonperturbative effects, as in \cite{Bachas:1999um}.

 The leading $\alpha'$-order terms of the  disk-level S-matrix elements, in general, have  both   contact terms as well as   massless poles. The contact terms along which produce the low energy effective  actions, may not satisfy the S-dual Ward identity. In fact, the low energy effective action of a single D$_3$-brane  which is given by the combination of the abelian Born-Infeld action \cite{Leigh:1989jq,Bachas:1995kx}  and the Chern-Simons action \cite{Polchinski:1995mt,Douglas:1995bn}, is ${\it not}$  invariant under the S-duality. However, it is known that their equations of motion   are invariant under the S-duality \cite{Gibbons:1995ap,Tseytlin:1996it,Green:1996qg}.

In general, an effective action can be separated into two parts, \ie
\beqa
S&=&S_1+S_2
\eeqa
where $S_1$ includes the couplings which are invariant under the linear S-duality, and $S_2$ includes the couplings which must be combined, in the momentum space, with some massless poles to become invariant under the linear S-duality. We will show that Born-Infeld action has such structure. That is, its $F^4$ terms are invariant under the linear  S-duality, however, its $F^6$ terms and higher  may be combined with some massless poles to become invariant under the linear S-duality. We will discuss how this property may be used iteratively to find all couplings of the effective action of D$_3$-branes.

The outline of the paper is as follows: In section 2, using the standard S-duality transformations at the linear order, we show that the S-matrix elements of one massless open string and one massless closed string vertex operators on the world volume of a D$_3$-brane can be written in a linear S-duality invariant form. In section 3 we show that the leading $\alpha'$-order terms of the disk-level S-matrix element of two  open string  and one closed string states can be combined into linear S-duality invariant combinations. Following \cite{ Bachas:1999um,Basu:2008gt}, the background dilaton factors in the nonleading order terms may be extended to the nonlinear $SL(2,Z)$ invariant functions. In section  4, we  show the same procedure can be applied for the S-matrix element of four abelian gauge field vertex operators.  We discuss our results in section 5.

\section{One gauge field  amplitudes}

In this section we consider the S-matrix elements of one gauge field and one closed string vertex operator. These amplitudes  in the string frame have been calculated in \cite{Garousi:1998fg}. In the Einstein frame, they are
\beqa
\cA(D_3;\z_1,B_2)&\sim&-T_3e^{-\phi_0}F_{1ab}B_2^{ab}\delta^{4}(k_1^a+p_2^a)\nonumber\\
\cA(D_3;\z_1,C^{(2)}_2)&\sim&\frac{T_3}{2}F_{1a_0a_1}C_{2a_2a_3}\epsilon^{a_0\cdots a_3}\delta^{4}(k_1^a+p_2^a)\labell{amp1}
\eeqa
where $\phi_0$ is the constant dilaton background, $\z_1$ is the polarization of the gauge field and $F_{1ab}=i(k_{1a}\z_{1b}-k_{1b}\z_{1a})$ is its field strength\footnote{Our index convention is that the Greek letters  $(\mu,\nu,\cdots)$ are  the indices of the space-time coordinates, the Latin letters $(a,d,c,\cdots)$ are the world-volume indices and the letters $(i,j,k,\cdots)$ are the normal bundle indices.}. $B_2$ and $C^{(2)}_2$ are the polarization of the B-field and the RR two-form, respectively. We have normalized the amplitudes such that they become consistent with T-duality. We have not, however, fixed the numerical factor of the amplitudes in this paper. To study the T-duality one should first change the Einstein frame metric $g^E_{\mu\nu}$ to the string frame metric $g^S_{\mu\nu}$ as $g^S_{\mu\nu}=e^{\phi_0/2}g^E_{\mu\nu}$, and then apply the linear T-duality transformation as in \cite{Garousi:2009dj,Garousi:2010rn}. In particular, the T-duality along a world volume direction maps the string frame tension  as $T_3\delta^{4}(k_1^a+p_2^a)\rightarrow T_2\delta^{3}(k_1^a+p_2^a)$ and the string frame coupling $F_{1ab}B_2^{ab}\rightarrow F_{1ab}B_2^{ab}+\cdots$ where dots refer to some couplings involving the transverse scalar fields in which we are not interested in this section. The D$_3$-brane tension in the string frame is $T_3=e^{-\phi_0}(2\pi)^{-3}(\alpha')^{-2}$ where $\alpha'$ is in the string frame. In the Einstein frame $\alpha'\rightarrow e^{-\phi_0/2}\alpha'$ and hence the tension becomes $T_3=(2\pi)^{-3}(\alpha')^{-2}$ which is invariant under the S-duality.

Now we have to show that the above amplitudes can be combined into a linear $SL(2,R)$ invariant form. 
The nonlinear transformation of the gauge field and the axion-dilaton, $\tau=C+ie^{-\phi}$,  are  given by \cite{Gibbons:1995ap}
\beqa
\left\{\matrix{F_{ab}&\rightarrow &sF_{ab}+r*G_{ab} \nonumber\\
G_{ab}&\rightarrow &pG_{ab}-q*F_{ab}}\right.\,\,\,;\,\,\,\tau\rightarrow \frac{p\tau+q}{r\tau+s}\labell{axidil}
\eeqa
where the antisymmetric tensor $G_{ab}$ is given by 
\beqa
G_{ab}&=&-\frac{2}{T_3}\frac{\prt L}{\prt F^{ab}}
\eeqa
where $L$ is the Lagrangian. 
Using $**=-1$, one can write the transformation of gauge field  as 
\beqa
\pmatrix{*F \cr 
G}\rightarrow (\Lambda^{-1})^T \pmatrix{*F \cr 
G}\,\,\,;\,\,\,\Lambda=\pmatrix{p&q \cr 
r&s}\labell{1}
\eeqa
The B-field and the RR two-form also appear as doublet under the $SL(2,R)$ transformation
\cite{Green:1996qg}
\beqa
\cB\equiv\pmatrix{B \cr 
C^{(2)}}\rightarrow (\Lambda^{-1})^T \pmatrix{B \cr 
C^{(2)}}\labell{2}
\eeqa
Unlike the transformations \reef{axidil}, the above transformation is  linear.
 
The S-matrix elements \reef{amp1} involve both the background fields as well as  the external open and closed string  fluctuations.  The invariance of the S-matrix elements under the linear S-duality is such  that the $SL(2,R)$ transformation on the background fields is nonlinear and on the external states is linear \cite{Garousi:2011we}. For the closed string amplitudes considered in \cite{Garousi:2011we}, this transforms an $n$-point function to another $n$-point function. In the cases that the S-matrix elements involve both open and closed strings, the transformation on the external states must be in such a way that an S-matrix element of  $n$-closed+$m$-open string vertex operators transforms to another S-matrix element of $n$-closed+$m$-open vertex operators. So the transformation of the open  and closed strings should be separately linear, \ie the open string should transform to open string, and the closed string should transform to closed string.  Since the gauge field in the  amplitudes \reef{amp1} is an open string quantum fluctuation, we have to consider vector to vector transformation on this field. To have linear  vector field in $G_{ab}$,  we have to consider the quadratic vector terms in the Lagrangian. Therefore, we have to consider the following  D$_3$-brane action in the Einstein frame  \cite{Gibbons:1995ap}:
\beqa
L=T_3\left(-\frac{1}{4}e^{-\phi_0}F_{ab}F^{ab}+\frac{1}{4}C_0F_{ab}(*F)^{ab}\right) \labell{L}
\eeqa
where  $(*F)^{ab}=\eps^{abcd}F_{cd}/2$ and $\phi_0,\, C_0$ are the background fields.  Note that there is no higher derivative corrections to the quadratic terms in \reef{L}. The antisymmetric tensor $G_{ab}$  becomes
\beqa
G_{ab}=e^{-\phi_0}F_{ab}-C_0(*F)_{ab}
\eeqa
which is linear in the vector field and has no higher derivative corrections at this order.

Now,  considering the transformation \reef{2} for the closed string fields and the following linearized transformation:
\beqa
\cF_{ab}\equiv\pmatrix{(*F)_{ab} \cr 
e^{-\phi_0}F_{ab}-C_0(*F)_{ab}}\rightarrow (\Lambda^{-1})^T \pmatrix{(*F)_{ab} \cr 
e^{-\phi_0}F_{ab}-C_0(*F)_{ab}}\labell{3}
\eeqa
for the open string field,
  the amplitudes \reef{amp1} can  be extended to 
\beqa
\cA(D_3;\z_1,\cB_2)&\sim&T_3(\cF_1^T)_{ab}\cN(\cB_2)^{ab}\delta^{4}(k_1^a+p_2^a)\labell{amp2}
\eeqa
where $\cF_1$ and  $\cB_2$ are the polarizations of the  $\cF$-field and the $\cB$-field, respectively. The 
 $SL(2,R)$ matrix $\cN$ is
\beqa
\cN=\pmatrix{0&1 \cr 
-1&0}
\eeqa
which  has the property 
\beqa
\cN=\Lambda {\cal N}\Lambda ^T
\eeqa
The  amplitude \reef{amp2} is manifestly invariant under the linear $SL(2,R)$ transformation. There is no background dilaton factor left over in \reef{amp2}, hence, there would be no need to include  loops and nonperturbative effects to make the amplitude fully invariant under the S-duality. The tree-level S-matrix element of 1-closed+1-open string vertex operators is the only S-matrix element which is fully invariant under the linear S-duality. In all other cases, one needs to add the loops and the nonperturbative effects  to make the tree-level amplitudes fully invariant under the S-duality. The observation that the amplitude \reef{amp2} is invariant under the S-duality is  consistent with the fact that  the loops effects in the 1-closed+1-open amplitude are zero.

 The S-dual amplitude \reef{amp2} includes the S-matrix elements \reef{amp1} as well as the following 2-point function in the presence of constant axion:
 \beqa
 \cA(D_3, C_0;\z_1,B_2)&\sim&\frac{T_3}{2}C_0F_{1a_0a_1}B_{2a_2a_3}\eps^{a_0\cdots a_3}\delta^{4}(k_1^a+p_2^a)\labell{amp3}
 \eeqa
 which is a standard coupling in the Chern-Simons part of the D$_3$-brane action. This S-matrix element can be calculated with disk-level 3-point function of one gauge field, one B-field and one RR scalar vertex operators in which the RR scalar is a constant. For non-constant RR field, the amplitude has the complicated structure of the S-matrix element of two closed and one open string vertex operators \cite{Fotopoulos:2001pt}, however, for constant field it should be reduced to \reef{amp3}. The disk-level 3-point function of one gauge field, one B-field and one RR  vertex operators has been recently calculated in \cite{Becker:2011ar}. It is easy to verify that for constant RR scalar, it reduces to \reef{amp3}.

\section{Two gauge fields  amplitudes}

The S-matrix element of two gauge fields and one closed string is nonzero when the closed string is dilaton, RR scalar, or graviton \cite{Hashimoto:1996kf,Garousi:1998fg}. Since graviton is invariant under the S-duality, one expects the S-matrix element of one graviton and two gauge fields to be invariant under the linear S-duality. On the other hand, the dilaton and the RR scalar transform as \reef{axidil}.
 So we expect the S-matrix elements of the dilaton and the axion combine into a linear S-dual multiplet. Let us first consider  the graviton amplitude.

The S-matrix element of two gauge fields and one graviton  is given in \cite{Hashimoto:1996kf,Garousi:1998fg}. In the Einstein frame, it is
\beqa
\cA(D_3;\z_1,\z_2,h_3)\sim T_3e^{-\phi_0}\left(F_{1a}{}^cF_{2bc}h_{3}^{ab}-\frac{1}{4}F_{1ab}F_2^{ab}h_{3c}{}^c\right)\frac{\Gamma(1-2te^{-\phi_0/2})}{[\Gamma(1-te^{-\phi_0/2})]^2}\labell{amp4}
\eeqa
where $h_3$ is the polarization of the graviton and the Mandelstam variable $t$  is $t=-\alpha' k_1\inn k_2$. There is also a conservation of momentum factor $\delta^{4}(k_1^a+k_2^a+p_3^a)$. Here again we have normalized the amplitude such that it becomes  consistent with linear T-duality. The $\alpha'$ in the Mandelstam variable $t$ which is in the Einstein frame, is invariant under the S-duality, however,  the dilaton factor in the Gamma functions in \reef{amp4} is not invariant under the S-duality. Hence,  we have to expand the Gamma functions in order to study the S-duality of this amplitude. The $\alpha'$ expansion of the Gamma functions is 
\beqa
\frac{\Gamma(1-2te^{-\phi_0/2})}{[\Gamma(1-te^{-\phi_0/2})]^2}=1+t^2\z(2)e^{-\phi_0}+2t^3\z(3)e^{-3\phi_0/2}+\frac{19}{4}t^4\z(4)e^{-2\phi_0}+\cdots
\eeqa
The first term is invariant, hence, to show that  the leading $\alpha'$-order term of \reef{amp4} is  invariant under the linear S-duality one should be able to write the kinematic factor in \reef{amp4} in linear $SL(2,R)$ invariant form. To this end,  consider  the matrix ${\cal M}$ 
 \beqa
 {\cal M}=e^{\phi}\pmatrix{|\tau|^2&C \cr 
C&1}\labell{M}
\eeqa
which transforms under the $SL(2,R)$ transformation as\footnote{Note that the matrix $\cM$ here is the inverse of the matrix $\cM$ in \cite{Gibbons:1995ap}.}
\beqa
{\cal M}\rightarrow \Lambda {\cal M}\Lambda ^T
\eeqa
Using this matrix, one finds 
\beqa
(\cF_1^T)_a{}^c\cM_0\cF_{2bc}=e^{-\phi_0}[(*F_1)_a{}^c(*F_2)_{bc}+F_{1a}{}^cF_{2bc}]
\eeqa
where $\cM_0$ is the matrix $\cM$ for constant background fields $\phi_0$ and $C_0$. Using the identity
\beqa
\eps_{a}{}^{cde}\eps_{bc}{}^{fg}=-\eta_{ab}(\eta^{df}\eta^{eg}-\eta^{dg}\eta^{ef})+\delta_a^f(\delta_b^d\eta^{eg}-\delta_b^e\eta^{dg})-\delta_a^g(\delta_b^d\eta^{ef}-\delta_b^e\eta^{df})
\eeqa
one finds
\beqa
(\cF_1^T)_a{}^c\cM_0\cF_{2bc}=e^{-\phi_0}[-\frac{1}{2}F_{1cd}F_2^{cd}\eta_{ab}+F_{1a}{}^cF_{2bc}+F_{1b}{}^cF_{2ac}]\labell{F12}
\eeqa
Using the above relation, one observes that  the kinematic factor in \reef{amp4} is invariant under the linear S-duality\footnote{The kinematic factor of two abelian gauge fields and two transverse scalars can be read from the expansion of DBI action. In the Einstein frame it is
\beqa
K(\z_1,\z_2,\Phi_3,\Phi_4)&=&e^{-\phi_0}\left(F_{1a}{}^cF_{2bc}\Phi_3^{i,a}\Phi_4^{j,b}\eta_{ij}-\frac{1}{4}F_{1ab}F_2^{ab}\Phi_3^{i,c}\Phi_4^{j,d}\eta_{ij}\eta_{cd}\right)+(1\leftrightarrow 2)\nonumber
\eeqa
where commas denote partial differentiation in the momentum space. This is similar to the kinematic factor in \reef{amp4}. The transverse scalar fields are invariant under the S-duality. Using \reef{F12}, one can write this kinematic factor as
\beqa
K(\z_1,\z_2,\Phi_3,\Phi_4)&=&(\cF_1^T)_a{}^c\cM_0\cF_{2bc}\Phi_3^{i,a}\Phi_4^{j,b}\eta_{ij}\labell{foot}
\eeqa
which is manifestly invariant under the linear S-duality.}. 

Therefore,   the leading term of the amplitude \reef{amp4} which is $\alpha'^0$-order, is invariant under the linear S-duality. All other terms  are the higher derivative terms. Since the linear S-duality transformation \reef{3} has no higher derivative corrections, all the higher derivative terms of \reef{amp4} are the higher derivatives of the kinematic factor. Hence, they all are  invariant under the linear S-duality. The $\alpha'^2$-order term, however,  has the constant dilaton factor  $e^{-\phi_0}$ which is not 
invariant under the nonlinear S-duality. The terms with the higher order of $\alpha'$ have other dilaton factors. None of them  are  invariant under the S-duality. 
Since the background dilaton and axion transform nonlinearly as \reef{axidil} under the S-duality, one should extend each of the dilaton  factors  to  a function of both dilaton and axion to make them invariant under the S-duality. In this way, one can find the exact  dependence of the amplitude on the background dilaton and axion.   By  adding the one-loop and the D-instanton effects to the $\alpha'^2$-order term, which may be done  by replacing $e^{-\phi_0}$ with the regularized non-holomorphi Eisenstein series $E_1(\phi_0,C_0)$, one may extend the $\alpha'^2$-order term  to the S-dual invariant form \cite{Bachas:1999um,Basu:2008gt,Garousi:2011fc}.  The dilaton factor $\z(3)e^{-3\phi_0/2}$ in the $\alpha'^3$-order term may be extended to the  non-holomorphic Eisenstein series $E_{3/2}(\phi_0,C_0)$ \cite{Basu:2008gt}. In general, the dilaton factor $\z(n)e^{-n\phi_0/2}$ may be extended to  the  non-holomorphi Eisenstein series $E_{n/2}(\phi_0,C_0)$ after including the loops and the nonperturbative effects. 

Therefore, the amplitude \reef{amp4} may be extended to 
\beqa
\cA(D_3;\z_1,\z_2,h_3)\sim \frac{T_3}{2}(\cF_1^T)_a{}^c\cM_0\cF_{2bc}h_{3}^{ab}\left(1+\alpha_2t^2E_{1}(\phi_0,C_0)+\alpha_3t^3E_{3/2}(\phi_0,C_0)+\cdots\right)\labell{amp5}
\eeqa
where $\alpha_n$s are some number, \ie $\alpha_2=1$, $\alpha_3=2$, and so on. This amplitude is invariant under the linear $SL(2,R)$ transformation on the external states and is invariant under the nonlinear $SL(2,Z)$ transformation on the background fields. One may expect the replacement of the dilaton factors in the tree-level amplitude with the appropriate   non-holomorphi Eisenstein series   includes all the loops and the noperturbative corrections to the tree-level amplitude, however, this does not mean that an exact S-matrix element can be found in this way. In general, there are many new terms in the loop amplitudes which have structure different than those in the tree-level. We expect the dilaton factors in the new terms at each loop order to become invariant under the S-duality after including the higher loops effects.

Now consider the S-matrix elements of dilaton and axion which  are given in \cite{Hashimoto:1996kf,Garousi:1998fg}. In the Einstein frame they are
\beqa
\cA(D_3;\z_1,\z_2,\phi_3)&\sim& T_3e^{-\phi_0}F_{1ab}F_2^{ab}\phi_3\frac{\Gamma(1-2te^{-\phi_0/2})}{[\Gamma(1-te^{-\phi_0/2})]^2}\nonumber\\
\cA(D_3;\z_1,\z_2,C_3)&\sim &T_3F_{1ab}(*F_2)^{ab}C_3\frac{\Gamma(1-2te^{-\phi_0/2})}{[\Gamma(1-te^{-\phi_0/2})]^2}\labell{amp6}
\eeqa
where $\phi_3$ is the polarization of dilaton, and  $C_3$ is the polarization of the axion. These polarizations are one, however, for clarity we keep them in the amplitudes. There is also a conservation of momentum factor $\delta^{4}(k_1^a+k_2^a+p_3^a)$ in each amplitude.

To write the kinematic factors in \reef{amp6} in a linear S-dual form, consider the variation of the matrix $\cM$ in \reef{M}. It is given by
\beqa
\delta\cM=\pmatrix{-(e^{-\phi}-C^2e^{\phi})\delta\phi+2C e^{\phi}\delta C& C e^{\phi}\delta\phi+e^{\phi}\delta C\cr 
C e^{\phi}\delta\phi+e^{\phi}\delta C&e^{\phi}\delta \phi}\labell{dM}
\eeqa
This matrix transforms under the $SL(2,R)$ transformation as
\beqa
{\delta\cal M}\rightarrow \Lambda \delta{\cal M}\Lambda ^T
\eeqa
Consider  the case that the variations are the external states, \ie $\delta\phi=\phi_3$ and $\delta C=C_3$, and the dilaton and the axion are the constant background fields $\phi_0$ and $C_0$, respectively. We call this matrix  $\delta\cM_3$. Using this matrix, one finds the following relation:
\beqa
(\cF_1^T)_{ab}\delta\cM_3\cF_2^{ab}=2e^{-\phi_0}F_{1ab}F_2^{ab}\phi_3+2F_{1ab}(*F_2)^{ab}C_3
\eeqa
which is invariant under the linear S-duality. Using this relation, one can write the kinematic factor in $\cA(D_3;\z_1,\z_2,\phi_3)+\cA(D_3;\z_1,\z_2,C_3)$  in linear $SL(2,R)$ invariant form. Adding the loops and the nonperturbative effects as in the previous case, one may   extend the dilaton factors in the $\alpha'$-expansion of the Gamma functions to the $SL(2,Z)$ invariant non-holomorphi Eisenstein series. Therefore, one may write the amplitudes \reef{amp6} as
\beqa
\cA(D_3;\z_1,\z_2,\phi_3+C_3)\sim \frac{T_3}{2}(\cF_1^T)_{ab}\delta\cM_3\cF_2^{ab}\left(1+\alpha_2t^2E_{1}(\phi_0,C_0)+\alpha_3t^3E_{3/2}(\phi_0,C_0)+\cdots\right)\labell{amp7}
\eeqa
which is manifestly invariant under the linear $SL(2,R)$ transformation on the quantum fluctuations and is invariant under the nonlinear $SL(2,Z)$ transformations on the background fields $\phi_0,\, C_0$.

\section{Four gauge fields   amplitude}

The  disk-level scattering amplitude of four gauge bosons on the world volume of a single D$_3$-brane  and for $1234$ ordering of the vertex operators is calculated in  \cite{Schwarz:1982jn}. In the Einstein frame, it is
\beqa
{\cal A}_{1234}&\sim& T_3\alpha'^2K(\z_1,\z_2,\z_3,\z_4)\frac{\Gamma(-se^{-\phi_0/2})\Gamma(-te^{-\phi_0/2})}{\Gamma(1-se^{-\phi_0/2}-te^{-\phi_0/2})}\delta^{4}(k_1^a+k_2^a+k_3^a+k_4^a)
\eeqa
where $s=-2\alpha'k_1\inn k_2$, $t=-2\alpha'k_1\inn k_4$ and the kinematic factor is \cite{Schwarz:1982jn}
\beqa
K&=&-e^{-2\phi_0}k_1\inn k_2(\z_1\inn k_4\z_3\inn k_2\z_2\inn\z_4+\z_2\inn k_3\z_4\inn k_1\z_1\inn\z_3+\z_1\inn k_3\z_4\inn k_2\z_2\inn\z_3+\z_2\inn k_4\z_3\inn k_1\z_1\inn\z_4)\nonumber\\
&&-e^{-2\phi_0}k_2\inn k_3 k_2\inn k_4 \z_1\inn \z_2\z_3\inn\z_4+\{1,2,3,4\rightarrow 1,3,2,4\}+\{1,2,3,4\rightarrow 1,4,3,2\}\labell{kin}
\eeqa
This kinematic factor is symmetric under any permutation of the external states and satisfies  the Ward identity associated with the gauge transformation.  The $\alpha'$-expansion of the Gamma functions is 
\beqa
\frac{\Gamma(-se^{-\phi_0/2})\Gamma(-te^{-\phi_0/2})}{\Gamma(1-se^{-\phi_0/2}-te^{-\phi_0/2})}&=&\frac{e^{\phi_0}}{st}-\frac{\pi^2}{6}-\z(3)(s+t)e^{-\phi_0/2}-\frac{\pi^4}{360}(4s^2+st+4t^2)e^{-\phi_0}+\cdots\nonumber
\eeqa
The total amplitude includes all non-cyclic permutation of the external states, \ie
\beqa
\cA=\cA_{1234}+\cA_{1243}+\cA_{1324}+\cA_{1342}+\cA_{1423}+\cA_{1432}
\eeqa
The $\alpha'$-expansion of the amplitude $\cA$ can be written as
\beqa
{\cal A}&\sim& T_3\alpha'^2K(\z_1,\z_2,\z_3,\z_4)\delta^{4}(k_1^a+k_2^a+k_3^a+k_4^a)\sum_{n=-2}^{\infty}a^{(n)}\labell{amp8}
\eeqa
where $a^{(n)}$s are functions of the Mandelstam variables. For abelian case in which we are interested, these functions are \cite{Bilal:2001hb}
\beqa
&&a^{(-2)}=a^{(-1)}=0\,,\,a^{(0)}=-\pi^2\,,a^{(1)}=0\,,a^{(2)}=-\frac{\pi^2}{4}(t^2+s^2+u^2)\z(2)e^{-\phi_0} \nonumber\\
&&a^{(3)}=-\pi^2stu \z(3)e^{-3\phi_0/2}\,,a^{(4)}=-\frac{9\pi^2}{48}(s^2+t^2+u^2)^2\z(4)e^{-2\phi_0}\,,\cdots
\eeqa
where $s+t+u=0$. 

To show that the amplitude satisfies the Ward identity corresponding to the global S-duality transformations, we first use the observation  in \cite{Gross:1986iv} that  indicates the leading contact terms of  $\cA$ are reproduced by the quartic terms of the BI action, \ie   $\Tr(F^4)/8-(\Tr(F^2))^2/32$ where the traces are over the world volume of the matrix $F_{ab}$. Using this observation, one can rewrite the kinematic factor in \reef{amp8}  as
 \beqa
 K=\cK_{1234}+\cK_{1243}+\cK_{1324}+\cK_{1342}+\cK_{1423}+\cK_{1432}
\eeqa
where
\beqa
\cK_{1234}=e^{-2\phi_0}\bigg[\frac{1}{2}\Tr(F_1F_2F_3F_4)-\frac{1}{16}\Tr(F_1F_2)\Tr(F_3F_4)-\frac{1}{16}\Tr(F_4F_1)\Tr(F_2F_3)\bigg]
\eeqa
Then using the relation \reef{F12}, one can write $K$ in the following simple form:
\beqa
K=\frac{1}{4}\bigg[\Tr(\cF_1^T\cM_0\cF_2\cF_3^T\cM_0\cF_4)+\Tr(\cF_1^T\cM_0\cF_3\cF_2^T\cM_0\cF_4)+\Tr(\cF_1^T\cM_0\cF_4\cF_2^T\cM_0\cF_3)\bigg]
\eeqa
which is invariant under the linear $SL(2,R)$ transformation.  Therefore, the amplitude \reef{amp8} at order $\alpha'^2$ is invariant under the linear $SL(2,R)$ transformation. Apart from the dilaton factors, all higher order terms are the derivative of the leading order terms which are then invariant under the linear S-duality. The dilaton factor $\z(2)e^{-\phi_0}$ in the $\alpha'^4$-order terms may be extended to the  regularized non-holomorphi Eisenstein series $E_1(\phi_0,C_0)$ after including the loop and the nonperturbative effects \cite{Green:1997tv,Bachas:1999um}. The dilaton factor $\z(3)e^{-3\phi_0/2}$ in the $\alpha'^5$-order terms may be extended to the  non-holomorphi Eisenstein series $E_{3/2}(\phi_0,C_0)$ \cite{Basu:2008gt}, and so on. 

The $\alpha'$-expansion of the S-matrix element of two gauge fields and two scalars is the same as \reef{amp8}, \ie 
\beqa
{\cal A}&\sim& T_3\alpha'^2K(\z_1,\z_2,\Phi_3,\Phi_4)\delta^{4}(k_1^a+k_2^a+k_3^a+k_4^a)\sum_{n=-2}^{\infty}a^{(n)}\labell{amp81}
\eeqa
The kinematic factor is the same as the four gauge boson case \reef{kin} in which the condition $\Phi\inn k=0$ is imposed and extra factor of $e^{\phi_0}$ is added because the indices of the scalars is uppercase, \ie $\Phi^i$, whereases the indices of the gauge fields is lowercase, \ie $A_a$.  It is shown in \reef{foot} that the kinematic factor can be written in linear $SL(2,R)$ invariant form. So the amplitude can be extended to S-dual form by including the loops and nonperturbative effects  as in the case of four gauge bosons.

Finally, the $\alpha'$-expansion of the S-matrix element of  four scalars is the same as \reef{amp8}, \ie 
\beqa
{\cal A}&\sim& T_3\alpha'^2K(\Phi_1,\Phi_2,\Phi_3,\Phi_4)\delta^{4}(k_1^a+k_2^a+k_3^a+k_4^a)\sum_{n=-2}^{\infty}a^{(n)}\labell{amp82}
\eeqa
The kinematic factor is the same as the four gauge boson case  \reef{kin} in which the condition $\Phi\inn k=0$ is imposed and extra factor of $e^{2\phi_0}$ is added.  So there is no dilaton  in the kinematic factor. The scalars are invariant under the S-duality\footnote{The invariance of the transverse scalar fields under the S-duality for the abelian case that we are interested in this paper, is as expected because there is no way to construct the  combination of the scalars to have either two antisymmetric world volume indices to contract with the world volume form under the S-duality, or three antisymmetric traverse indices to contract with the transverse volume form under the S-duality. For the nonabelian case, however, the second possibility can occur which has been considered in \cite{Taylor:1999pr}. }, hence, the kinematic factor is invariant under the S-duality. The amplitude can be extended to the $SL(2,Z)$  invariant form  as in the case of four gauge fields. This ends our illustration of consistency of the S-matrix elements on the world volume of abelian D$_3$-brane with the Ward identity corresponding to the global S-duality transformations.

\section{Discussion}

In this paper, by working on some examples, we have shown  that the S-matrix elements of n-closed+m-open string vertex operators on the world volume of a single D$_3$-brane at the leading order of $\alpha'$ can be combined into multiplets which are invariant under the linear $SL(2,R)$ transformations. The extra  dilaton factors in the nonleading order terms may become  invariant under S-duality after including the loops and the nonperturbative effects \cite{Green:1997tv,Bachas:1999um}. 

The S-duality transformations on the  quantum fluctuations (classical fields) must be linear (nonlinear) \cite{Garousi:2011we}. Moreover, since the S-matrix element of  $n$ open strings is totally different from the S-matrix element of $n$  closed strings, the linear S-duality should  transform  open string  to open string and closed string to closed string.   In all examples we have worked out, the leading $\alpha'$ order terms of the multiplets are invariant under the above linear $SL(2,R)$ transformations.

The D$_3$-brane is invariant under the S-duality, so the fundamental string excitation of D$_3$-brane at weak coupling should transform to the (p,q)-string excitation of  the D$_3$-brane under the $SL(2,Z)$ transformation. Moreover, the external string should transform to the (p,q)-string as well. The massless spectrum of these strings are invariant under the linear S-duality, \eg the   $C^{(2)}$ of the fundamental closed string at weak coupling maps to $B^{(2)}$ of the closed D-string  at strong coupling, or the electric components of the gauge field strength $F$ of the fundamental open string at weak coupling maps to the magnetic components of the gauge field strength of  the open D-string  at strong coupling. 

In general, one expects an S-matrix element at weak coupling transforms under the linear S-duality to another S-matrix element at strong coupling which is different from the original one. Consider, for example,  the weak coupling S-matrix element \reef{amp4}  which is at zero axion background and has  two external massless open and one massless closed fundamental strings. This amplitude transforms under the linear S-duality to the S-matrix element of  two  open and one  closed D-string at strong coupling.  At zero axion background, it  is given by 
\beqa
\cA(D_3;\z_1,\z_2,h_3)\sim T_3e^{-\phi_0}\left(F_{1a}{}^cF_{2bc}h_{3}^{ab}-\frac{1}{4}F_{1ab}F_2^{ab}h_{3c}{}^c\right)\frac{\Gamma(1-2te^{\phi_0/2})}{[\Gamma(1-te^{\phi_0/2})]^2}\labell{amp41}
\eeqa
where now the gauge fields and the graviton are the massless modes of the external D-strings. The poles of  the Gamma functions show   the open D-string excitation of the D$_3$-brane \cite{Garousi:2011we}. The above amplitude can not be calculated in the perturbative string theory. The kinematic factors in the two amplitudes are the same, however, the Gamma functions are different as expected. The difference between the above amplitude and the amplitude \reef{amp4}  steams from the fact that the original amplitude \reef{amp4} does not include the loops and the nonperturbative effects. Obviously, if one includes these effects which may be given by \reef{amp5}, then the transformed amplitude would be the same as the original one.


The linear S-duality invariance  of the leading $\alpha'$ order terms of the S-matrix elements indicates that the low energy effective field theory on the world volume of D$_3$-brane may not be invariant under the linear S-duality. To see this consider the following discussion:
The leading $\alpha'$-order terms of the  S-matrix element \reef{amp8} are reproduced by the $F^4$ terms of the BI action \cite{Gross:1986iv}. Hence, the $F^4$ terms of BI action are invariant under the linear S-duality transformations. On the other hand,  up to a total derivative term the $F^2$ term of this action which is not invariant under the S-duality is zero when the gauge field is on-shell. So the on-shell BI action is invariant under the linear S-duality up to $F^4$ terms. What happens for the $F^6$ and the higher order terms? Are they invariant under the linear S-duality as well? Consider the expansion of  the BI action in the Einstein frame, \ie
\beqa
\sqrt{-\det(\eta_{ab}+e^{-\phi_0/2}F_{ab})}-1&=&-\frac{e^{-\phi_0}}{4}\Tr(F^2)-\frac{e^{-2\phi_0}}{8}\left[\Tr(F^4)-\frac{1}{4}(\Tr(F^2))^2\right]\labell{BI}\\
&&-\frac{e^{-3\phi_0}}{12}\left[\Tr(F^6)-\frac{3}{8}\Tr(F^2)\Tr(F^4)+\frac{1}{32}(\Tr(F^2))^3\right]+\cdots\nonumber
\eeqa
One may use  the relations specific to four dimensions to write the above terms in alternative ways. For example, one can show the following relation for $F^6$ terms:
\beqa
\Tr(F^6)-\frac{3}{4}\Tr(F^2)\Tr(F^4)+\frac{1}{8}(\Tr(F^2))^3=0
\eeqa
Using this relation one may rewrite the $F^6$ terms in \reef{BI} in a different form.

Now,  using \reef{F12} one finds the following expressions are invariant under the linear $SL(2,R)$ transformation:
\beqa
0&=&\Tr(\cF^T\cM_0\cF)\nonumber\\
4e^{-2\phi_0}\bigg[\Tr(F^4)-\frac{1}{4}(\Tr(F^2))^2\bigg]&=&\Tr(\cF^T\cM_0\cF\cF^T\cM_0\cF)\labell{SBI}\\
8e^{-3\phi_0}\left[\Tr(F^6)-\frac{3}{4}\Tr(F^2)\Tr(F^4)+\frac{1}{8}(\Tr(F^2))^3\right]&=&\Tr(\cF^T\cM_0\cF\cF^T\cM_0\cF\cF^T\cM_0\cF)\nonumber
\eeqa
The first two terms are the same as the first two terms in the on-shell BI action \reef{BI}. However, the $F^6$ terms in \reef{SBI} are not the same as the corresponding terms in  \reef{BI}. In fact,  they add up to  zero. More generally, one can show that the couplings which are invariant under the linear S-duality are
\beqa
\Tr(\overbrace{\cF^T\cM_0\cF\cdots \cF^T\cM_0\cF}^{(2n-1)})&=&0\nonumber\\
\Tr(\overbrace{\cF^T\cM_0\cF\cdots \cF^T\cM_0\cF}^{(2n)})&=&4e^{-2n\phi_0}\bigg[\Tr(F^4)-\frac{1}{4}(\Tr(F^2))^2\bigg]^n\labell{rel2}
\eeqa
where in the first case the number of $\cF^T\cM_0\cF$ are odd and in the second case it is even. There are no such simple relations for the corresponding terms in \reef{BI}

The reason that the $F^6$ terms of the BI action are  not invariant under the  linear $SL(2,R)$ transformation is that the S-matrix element of the six gauge field vertex operators at leading order has both contact terms and massless poles. The contact terms are reproduced by the $F^6$ of the BI action \reef{BI}, and the massless poles are reproduced by the $F^4$ terms of \reef{BI}. 
The massless pole in $(k_1+k_2+k_3)^2$-channel, for example, is given by the following Feynman amplitude:
\beqa
{\cal A}_{{123}} &=&V^a(\z_1,\z_2,\z_3, A)G_{ab}(A)V^b(A,\z_4,\z_5,\z_6)\labell{pole}
\eeqa
where the propagator can be read from the $F^2$ term of \reef{BI}
and the vertexes can be read from the $F^4$ terms of \reef{BI}. 
Neither the $F^6$-massless-poles nor the $F^6$-contact-terms are invariant under the linear S-duality. According to the S-duality proposal,  the combination of these two terms, \ie ($F^6$-massless-poles + $F^6$-contact-terms), must  be invariant under the linear S-duality\footnote{It has been shown in \cite{Rosly:2002jt,Boels:2008fc} that the tree-level scattering amplitudes generated by BI action conserve helicity which  might be a result of   the S-duality.}.  Similar observation has been made in \cite{Garousi:2010rn} in trying to extend the anomalous Chern-Simons couplings at order $\alpha'^2$ to be consistent with the linear T-duality. In that case also one observes that only the combination of the contact terms and massless poles are fully invariant under the T-duality.


 One may find the $F^{2n}$  terms of the BI action by imposing the linear S-duality of ($F^{2n}$-massless-poles + $F^{2n}$-contact-terms). Starting from $F^4$ terms which are invariant under the linear S-duality, one can construct $F^{6}$-massless-poles. Then imposing the linear S-duality, one may find  $F^{6}$-contact-terms. Using the $F^4$- and the $F^6$-contact terms, one can calculate the $F^{8}$-massless-poles. Then using the linear S-duality, one may find the $F^{8}$-contact-terms, and so on. This calculation should  confirms  the abelian BI action as the effective action of a single D-brane. 
 
 When there are more than one coincident D-branes, the abelian BI  action should be extended to a nonabelian gauge theory. A nonabelian extension of the BI action has been proposed in \cite{Tseytlin:1997csa,Tseytlin:1999dj} which includes the symmetric trace prescription. This proposal works for $F^2$ and $F^4$-terms, however, there are various discussions that indicate the $F^6,\, F^8, \, \cdots$ terms of the D-branes are not correctly captured by the nonabelian BI action \cite{Hashimoto:1997gm,Bain:1999hu}.  One may extended the above linear S-duality of ($F^{2n}$-massless-poles + $F^{2n}$-contact-terms) to  the nonabelian case to find the $F^n$-terms of the nonabelian D$_3$-branes action.   It would be interesting then to study the Ward identity corresponding to the global  S-duality transformations  for the nonabelian cases.

 {\bf Acknowledgments}:   This work is supported by Ferdowsi University of Mashhad under grant 2/18717-1390/07/12.



\end{document}